\documentclass[12pt]{article}%
\usepackage{graphicx}

\begin{document}
 
\begin{center}{\LARGE\bf Finite mathematics as the most general (fundamental) mathematics} \end{center}

\vskip 1em \begin{center} {\large Felix M. Lev} \end{center}

\begin{center} Email:  felixlev314@gmail.com \end{center}
\begin{abstract}
The purpose of this paper is to explain at the simplest possible level
why finite mathematics based on a finite ring of characteristic $p$ is more general (fundamental) than standard mathematics. The belief of most mathematicians and physicists that standard mathematics is the most fundamental arose for historical reasons. However, simple {\it mathematical} arguments show that standard mathematics (involving the concept of infinities) is a degenerate case of finite mathematics in the formal limit $p\to\infty$: standard mathematics arises from finite mathematics in the degenerate case when operations modulo a number are discarded. Quantum theory based on a finite ring of characteristic $p$ is more general than
standard quantum theory because the latter is a degenerate case of the former in the formal limit $p\to\infty$.
\end{abstract}

\begin{flushleft} Keywords: finite mathematics, standard mathematics, finite quantum theory\end{flushleft}
\begin{flushleft} MSC 2020: 11Axx, 11Txx, 13Mxx, 16Gxx, 81R05\end{flushleft}

\begin{center} {\bf  List of Abbreviations} \end{center}
\begin{flushleft} FM: finite mathematics\end{flushleft}
\begin{flushleft} SM: standard mathematics\end{flushleft}
\begin{flushleft} SR: special relativity \end{flushleft}
\begin{flushleft} NM: nonrelativistic mechanics\end{flushleft}
\begin{flushleft} QT: quantum theory\end{flushleft}
\begin{flushleft} CT: classical theory\end{flushleft}
\begin{flushleft}  FQT: Quantum theory based on finite mathematics \end{flushleft}
\begin{flushleft}  SQT: Standard quantum theory \end{flushleft}
\begin{flushleft} IR: irreducible representation\end{flushleft}
\begin{flushleft}  QFT: Quantum Field Theory\end{flushleft}
\begin{flushleft}  NQT: Nonrelativistic Quantum Theory\end{flushleft}
\begin{flushleft}  RQT: Relativistic Quantum Theory\end{flushleft}
\begin{flushleft}  dS: de Sitter \end{flushleft}
\begin{flushleft}  AdS: Anti de Sitter \end{flushleft}
\begin{flushleft}  dSQT: de Sitter Quantum Theory \end{flushleft}
\begin{flushleft}  AdSQT: Anti de Sitter Quantum Theory \end{flushleft}

\section{The main goal of this paper}
\label{goal}

In \cite{book,arxiv} and other our publications we investigated in detail why
finite mathematics based on a finite ring of characteristic $p$ is more general (fundamental) than standard mathematics. These publications contain detailed proofs of statements on which
our approach is based. The purpose of this paper is to explain the main ideas of our approach
at the simplest possible level. Therefore, we do not provide technical details of the proofs but for interested readers we provide links how those proofs can be found.

SM deals with relations
\begin{equation}
a+b=c,\,\, a\cdot b=c,\,\, etc.	
\label{standard}
\end{equation}
On the other hand, FM deals with relations
\begin{equation}
	a+b=c\,\,(mod\, p),\,\, a\cdot b=c\,\,(mod\, p),\,\, etc.	
\label{finite}
\end{equation}
where all the numbers $a,b,c,...$ can take only values $0,1,2,...p-1$ and $p$ is called characteristic of the ring. Therefore, in FM there are no infinities and all numbers do not exceed $p$ in absolute value. 

Before discussing these versions of mathematics, let's discuss the following. How should we treat mathematics: i) as a purely abstract science or ii) as a science that should describe nature? I am a physicist and have worked among physicists for most of my life. For them, only approach ii) is acceptable. However, when I discussed this issue with mathematicians and philosophers, I discovered that many of them view mathematics only from the point of view of i) and arguments related to the description of nature are not significant for them. 
Approach i) can be called the approach of Hilbert, who was its most famous proponent.
There is a great discussion in the literature between him and Gödel about whether Gödel's incompleteness theorems indicate that the approach has foundational problems.

The fact that Hilbert's approach does not raise the question of describing nature does not mean that this approach should be rejected out of hand. For example, Dirac's philosophy is: {\it "I learned to distrust all physical
concepts as a basis for a theory. Instead one should put one’s trust in a mathematical scheme, even if the scheme does not appear at first sight to be connected with physics. One should concentrate on getting an interesting mathematics."} Dirac also said that for him the most important thing in any physical theory is the beauty of formulas in this theory. That is, he meant that sooner or later, in any beautiful mathematical theory, its physical meaning will be found. But even if it is not found, the beauty of the theory itself has aesthetic value. For example, in music we appreciate its beauty and do not demand that music should somehow describe nature.

Nevertheless, in this paper, we treat mathematics only as a tool for describing nature. 
In the framework of this approach, most mathematicians and physicists believe that, at the most fundamental level, nature is described by SM, and FM is needed only in some special model problems. This opinion has developed despite the fact that modern quantum theory has known problems and, despite the numerous efforts of many highly qualified mathematicians and physicists over the years, these problems have not yet been solved. 

Modern QFT can calculate observable quantities only within the framework of perturbation theory, and it is not known whether its series is convergent or only asymptotic. However, even within this framework, one of the key problems of QFT (based on SM) is the problem of divergences: the theory gives divergent expressions for the S-matrix. In renormalized theories, the divergences can be eliminated by renormalization where finite observable quantities are formally expressed as products and sums of singularities. From the mathematical point of view, such procedures are not legitimate but in some cases they result in impressive agreement with experiment. The most famous case is that the results for the electron and muon magnetic moments obtained at the end of 40th agree with experiment with the accuracy of eight decimal digits. In view of this and other
successes of QFT, most physicists believe that agreement with the data
is much more important than the rigorous mathematical substantiation.

At the same time, in non-renormalized QFTs, divergences cannot be eliminated by  renormalization, and this is a great obstacle for constructing quantum gravity based on QFT. As the famous Nobel Prize laureate Steven Weinberg wrote in his book \cite{Weinberg1}: "{\it Disappointingly this problem appeared with even greater severity in the early days of quantum theory, and although greatly ameliorated by subsequent improvements in the theory, it remains 	with us to the present day}". The title of Weinberg's paper \cite{Weinberg2} is "Living with infinities".

The main goal of the present paper is to explain at the simplest possible level that, contrary to the belief of most mathematicians and physicists, FM is the most general (fundamental) mathematics, and SM is its degenerate case. For this purpose it is necessary to give a definition  when mathematics A is more general (fundamental) than mathematics B, and mathematics B is a degenerate case of mathematics A. In \cite{book,arxiv} we have proposed the following 

{\bf Definition:} {\it Let theory A contain a finite nonzero parameter and theory B be obtained from theory A in the formal limit when the parameter goes to zero or infinity. Suppose that, with any desired accuracy, A can reproduce any result of B by choosing a value of the parameter. On the contrary, when, the limit is already
taken, one cannot return to A and reproduce all results of A. Then A is more general than B and B is a degenerate case of A}. 

In this paper we  discuss the result of \cite{book,arxiv} that, as follows from {\bf Definition}, contrary to the belief of most mathematicians and physicists:

{\bf Statement:} {\it SM is a degenerate
	case of FM in the formal limit $p\to\infty$, where $p$ is the characteristic of the ring in FM.}

As explained below, this {\bf Statement} implies that any result of SM can be obtained in FM with some choice of $p$, and, on the other hand, SM cannot reproduce those results of FM where it is important that $p$ is finite and not infinitely large. As explained below, a consequence of this {\bf Statement} is 
that FM is more general (fundamental) than SM because SM is obtained from FM in the case when all operations modulo a number are discarded. Also, as discussed in \cite{book,arxiv,lev3} and this paper, a consequence of this {\bf Statement} is that, {\it for describing nature at the most fundamental level, the concepts of infinitesimals, infinitely large, limits, continuity etc. are not needed; they are needed only for describing nature approximately}. 

Kronecker's famous phrase is that God invented integers, and humans invented everything else. However, in view of this {\bf Statement}, this phrase can be reformulated so that God came up with only finite sets of numbers, and everything else was invented by people.

As follows from {\bf Statement}, in QT based on FM (which we call finite quantum theory - FQT), the problem of divergences does not exist in principle because
in FM there are no infinities. We emphasize that {\bf Statement} is not only our wish, but a fact proven mathematically in \cite{book,arxiv,lev3} and Sec. \ref{finmath}. Therefore, those mathematicians and physicists who insist on their position that SM is more general (fundamental) than FM must either give arguments that {\bf Definition} is not justified or show that the proof in \cite{book,arxiv,lev3} and Sec. \ref{finmath} is erroneous. However, in numerous discussions with me, those mathematicians and physicists have presented various arguments that, in their opinion, emphasize the correctness of their position. Typical arguments are:
\begin{itemize}
	\item a) Formally, you have no divergences, but you introduce the cutoff $p$ which is a huge number. Therefore, in cases when infinities arose in the standard theory, you will get a huge number $p$ which is practically infinite.
	\item b) In your theory there is only one parameter $p$ and it is not clear why this parameter is this and not another. Is not it reasonable to prefer the approach with adeles when there are many characteristics which are on equal footing?
	\item c) An argument that has some similarities with b) is this: when you say that God only invented finite sets of numbers and everything else (infinitesimals, infinitely large etc.) was invented by people, do you think that he "invented" a biggest (finite) $p$? 
\end{itemize}
These arguments will be discussed below.

The paper is organized as follows. In Sec. \ref{finmath} we explain why a theory proceeding
from a finite ring is more general than a theory proceeding from the infinite ring $Z$.
In Sec. \ref{SR} we explain why special relativity where speeds cannot be greater than $c$ is
more general than classical mechanics where there is no speed limit. 
In Sec. \ref{FQTvsSQT} we describe the main ideas of quantum theory based on finite mathematics. In 
Secs. \ref{FQT} and \ref{examples} we explain why quantum theory based on finite mathematics is more general
(fundamental) than standard quantum theory. In section \ref{answers} we answer questions that are commonly asked in connection with our approach.

\section{Basic facts about finite mathematics}
\label{finmath}

SM starts from the infinite ring of integers $Z=(-\infty,...-1,0,1,...\infty)$ but FM can involve only a
finite number of elements. 
FM starts from the ring $R_p=(0, 1, 2, ... p-1)$ where addition, subtraction and multiplication are defined
as usual but modulo $p$. In our opinion, the notation $Z/p$ for $R_p$ is not adequate because it may
give a wrong impression that FM starts from the infinite set $Z$
and that $Z$ is more general than $R_p$. However, although $Z$ has more elements than $R_p$, $Z$ cannot 
be more general than $R_p$ because $Z$ does not contain operations modulo a number. If $p$ is prime then $R_p$ becomes
the Galois field $F_p$ where all the four operations are possible. The number $p$ is called the characteristic
of the ring $R_p$ or the field $F_p$. For example, if $p=5$ then 3+1=4 as usual but 3$\cdot$2=1, 4$\cdot$3=2, 4$\cdot$4=1 and 3+2=0. Therefore -2=3, -4=1, 1/2=3, 1/4=4 etc. The theory of finite rings and fields is described in standard textbooks (see e.g., \cite{VdW,Ireland,Davenport}).

One might say that the above examples have nothing to do with reality since 3+2 always equals 5 and not zero.
However, since operations in $R_p$ are modulo $p$, one can represent 
$R_p$ as a set $\{0,\pm 1,\pm 2,...,\pm(p-1)/2)\}$ if $p$ is odd or as a set
$\{0,\pm 1,\pm 2,...,\pm (p/2-1),p/2\}$ if $p$ is even. Let $f$ be a function from $R_p$ to $Z$ such that
$f(a)$ has the same notation in $Z$ as $a$ in $R_p$. Then for elements $a\in R_p$ such that $|f(a)|\ll p$, 
addition, subtraction and multiplication are the same as in $Z$. In other words, for such elements we do not notice the existence of $p$. 

One might say that nevertheless the field $F_p$ cannot be used in physics since 1/2=($p$+1)/2 i.e., a very large
number when $p$ is large. However, as explained in \cite{book,arxiv,lev3} and Sec. \ref{FQTvsSQT}, {\it since quantum states are projective then, even in SQT, quantum states can be described with any desired accuracy by using only integers and therefore the concepts of rational and real numbers play only an auxiliary role}.

\begin{figure}[!ht]
	\centerline{\scalebox{0.3}{\includegraphics{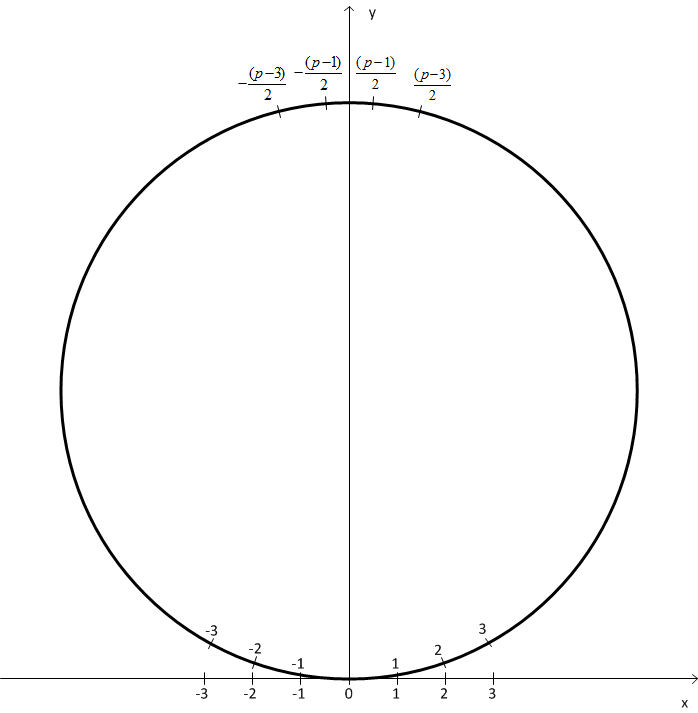}}}
	\caption{
		Relation between $R_p$ and $Z$
	}
	\label{Fig.2}
\end{figure}

If elements of $Z$ are depicted as integer points on the $x$ axis of the $xy$ plane then, if $p$ is odd, the elements of $R_p$
can be depicted  as points of the circumference in Figure \ref{Fig.2}
and analogously if $p$ is even.
This picture is natural from the following considerations. As explained in textbooks, both $R_p$ and $Z$ 
are cyclic groups with respect to addition. However, 
$R_p$ has a higher symmetry because it has a property which we call {\it strong cyclicity}:
if we take any element $a\in R_p$ and sequentially add 1 then after $p$ steps we will exhaust
the whole set $R_p$ by analogy with the property that if we move along a circumference in the same direction
then sooner or later we will arrive at the initial point. At the same time, if we take an element $a\in Z$ then
the set $Z$ can be exhausted only if we first successively add +1 to $a$ and then -1 to $a$ or {\it vice versa} and those
operations should be performed an infinite number of times.
As noted in \cite{book,arxiv}, in FQT strong cyclicity plays an important role. In particular, it explains why one IR of the symmetry algebra
describes a particle and its antiparticle simultaneously.

The above construction has a known historical analogy. For many years people
believed that the Earth was flat and infinite, and only after a long period of time
they realized that it was finite and curved. It is difficult to notice the curvature when
we deal only with distances much less than the radius of curvature. Analogously
one might think that the set of numbers describing physics in our universe has a
“curvature” defined by a very large number $p$ but we do not notice it when we deal
only with numbers much less than $p$.

By analogy with SM, one can say that if $a\in R_p$ then $a>0$ if $f(a)>0$ and $a<0$ if $f(a)<0$.
In other words, "positive" elements of $R_p$ are on the right half-circle of Figure 1 and
"negative" elements --- on the left half-circle. In SM, if $a>0$ and $b>0$ then $(a+b)>0$.
However, in FM this is not necessarily the case because the operations here are modulo $p$.
For example, $(p-1)/2>0$ and $1>0$ but $(p-1)/2+1=(p+1)/2=-(p-1)/2$, i.e., $f((p-1)/2+1)<0$.
Therefore, in $R_p$, the concepts of $>$ and $<$ have the same meaning as in SM only if
they apply to numbers $a$ such that $|f(a)|$ is much less than $p$.

In Sec. 6.3 of \cite{book} and \cite{arxiv}, it is proved that, as follows from {\bf Definition}, 

{\it Statement 1: The ring $R_p$ is more general than the ring $Z$
	and the latter is a degenerate case of the former in formal limit $p\to\infty$}.
	
This implies that the ring $Z$ is the limit of the ring $R_p$ when $p\to\infty$. 
Note that {\it  in the technique of SM, infinity is understood only as a limit (i.e., as  potential infinity)
but the basis of SM does involve actual infinity}: SM starts from the infinite ring $Z$ and, even in standard textbooks on mathematics, it is not even posed a problem whether $Z$ can be treated as a limit of finite rings.
The problem of actual infinity is discussed in a vast literature, 
and, in SM, $Z$ is treated as actual and not potential infinity, i.e., there is 
no rigorous definition of $Z$ as a limit of finite rings. Moreover, classical set theory considers
infinite sets with different cardinalities.

As explained in \cite{book,arxiv,lev3}, {\it Statement 1} is the basic stage in proving {\bf Statement}, i.e., that FM
is more general than SM. In particular, in the approach ii), this means that FQT is more general (fundamental) than SQT. This issue will be also discussed in Secs. \ref{FQTvsSQT} and  \ref{examples}. 
Therefore {\it Statement 1} should not be based on the
properties of the ring $Z$ derived in SM. 
The statement should be proved
by analogy with standard proof that a sequence of natural numbers
$(a_n)$ goes to infinity if $\forall M>0$ $\exists n_0$ such that $a_n\geq M\,\, \forall n\geq n_0$. In particular, the proof should involve only potential infinity but
not actual one.

The meaning of {\it Statement 1} is
that for any $p_0>0$ there exists a set $S$ belonging to all $R_p$ with $p\geq p_0$ and a natural number $n$ such that for any 
$m\leq n$ the result of any $m$ operations of summation, subtraction or multiplication of elements 
from $S$ is the same as in $R_p$ for any $p\geq p_0$ and that cardinality
of $S$ and the number $n$ formally go to infinity when $p_0\to\infty$. This means 
that for the set $S$ and number $n$ there is no
manifestation of operations modulo $p$, i.e., the results of any $m\leq n$ operations
of elements from $S$ are formally the same in $R_p$ and $Z$. This implies that for 
experiments involving only such sets $S$ and numbers $n$
it is not possible to conclude whether the experiments are described by a
theory involving $R_p$ with a large $p$ or by a theory involving $Z$.

As noted e.g.,
in \cite{book,arxiv}, the fact that $Z$ can be treated as a limit of $R_p$ when $p\to\infty$ follows from a construction called ultraproducts.
However, theory of ultraproducts  is essentially based on classical results involving actual infinity, in particular, on \L{}o\^s' theorem involving the
axiom of choice. Therefore theory of ultraproducts cannot be used in proving
that FM is more general than SM.

When the radius of the circumference in Figure \ref{Fig.2} becomes infinitely large then
a vicinity of zero in $R_p$ becomes the infinite set $Z$ when $p\to\infty$. 
Therefore {\it even from pure mathematical point of view, the concept of infinity cannot be fundamental because, as soon as we involve infinity and replace $R_p$ by $Z$, we automatically obtain a degenerate theory because in $Z$ there are no operations modulo a number}.

In FQT, states are elements of linear spaces over $R_p$. One might think that SQT is a more general theory than FQT
because in SM, $Z$ is generalized to the case of rational and real numbers. However, as noted 
in \cite{book,arxiv} and Sec. \ref{FQTvsSQT}, since in SQT the states are projective, then, {\it even in standard quantum theory}, it suffices to use only integers for describing experimental data with any desired accuracy.

\section{Analogy between SR and FM}
\label{SR}

As noted in Sec. \ref{goal}, in the standard physics literature, the fundamental nature of various physical theories is discussed on the basis of physical considerations. However, in Sec. \ref{FQTvsSQT}, we will discuss purely mathematical criteria 
for comparing the fundamentality of various physical theories over SM. Nevertheless, for illustrative purposes, in this section
we consider comparison of SR and NM from a point of view of a very simple example. 

Before the creation of SR, it was believed that NM was the most general (fundamental) mechanics. There are no restrictions on the magnitude of speed there which can be in the interval $[0,\infty )$. However, in SR, the speed cannot exceed $c$.

The fact that there is a speed limit greatly changes the standard philosophy of NM. For example, in NM it seems unnatural that the speed of $0.99c$ is possible, but $1.01c$ is not. For this and other reasons, it took a long time for SR to be accepted by the majority of physicists.

Let's consider a simple model example when in our reference frame some observer moves with speed $v_1$ and in the reference frame of this observer some particle moves in the same direction with speed $v_2$. Then, according to the rules of NM, the speed of the particle in our reference frame will be $V=v_1+v_2$. So, even if $v_1<c$ and $v_2<c$ then, in NM, a situation is possible when $V>c$ and this may suggest that the statement of SR about the speed limit is not consistent. However, the result of SR in such a situation is not $V=v_1+v_2$ but
\begin{equation}
V=\frac{v_1+v_2}{1+v_1v_2/c^2}
\end{equation}
and this value cannot exceed $c$. 
In particular, if $v_1=v_2=0.6c$ then $V$ is not equal to $1.2c$ as one might think from naive considerations, but $V\approx0.882c$, and if
$v_1=v_2=0.99c$ then $V$ is not equal to $1.98c$ but $V\approx 0.9999495c$. The lesson
of this example is that it is not always correct to make judgments proceeding from ''common sense".

Here there is an analogy with FM: for example, if $a$ and $b$ are such natural numbers that $a<p$, $b<p$ and in SM there may be a situation when $(a+b)>p$, then in FM such a situation cannot exist because always $(a+b)\,\, (mod\,\,p) <p$.

It is now generally accepted that SR is confirmed experimentally to a greater extent than NM. Also, as explained in Sec. \ref{FQTvsSQT}, it follows from {\bf Definition} that NM is a degenerate case of SR since SR can reproduce any fact of NM with some choice of $c$, while NM cannot reproduce those facts of SR in which it is essential that $c$ is finite and not infinite. Thus, SR does not disprove NM, but shows that it works with high accuracy when speeds are much less than $c$. There is an analogy here with the fact that, as shown in Sec. \ref{finmath}, FM does not refute SM, but shows that the latter is a good approximation to reality only in situations where the numbers in a given problem are much less than $p$.

In complete {\it logical} analogy with the objections to FM in points (a-c) in Sec. \ref{goal}, one can put forward similar objections to SR, but now the role of $p$ will be played by $c$. Therefore, I think that, {\it for being completely consistent, if we reject FM, we must also reject SR, and if we accept SR then, by the same logic, we must also accept that FM is more general (fundamental) than SM}.

As follows from the above results, it is not necessary to apply SR in everyday life
when speeds are much less than $c$ because in this case NM
works with a very high accuracy. 
Analogously, for describing almost all phenomena at the macroscopic level, there is no need to apply QT. For example,
there is no need to describe the motion of the Moon by the Schrödinger equation. In principle this is possible but results in unnecessary complications. At the same time, microscopic phenomena can be correctly described only in the
framework of QT. 

\section{Quantum theory based on finite mathematics}
\label{FQTvsSQT}

In QFT, symmetry at the quantum level is described as follows. First, the existence of some background space-time is postulated - Galilei, Minkowski, dS, AdS or some other background.  This background has a group of motions. It is postulated that the basic operators for the system under consideration commute as required in the Lie algebra representation of this group. That is, these operators form a representation of the Galilei algebra, Poincare algebra, dS algebra, AdS algebra or some other algebra. This approach to symmetry is in the spirit of Felix Klein's Erlangen program.

The Erlangen program was proposed in 1872 when quantum theory did not yet exist. As discussed in detail in \cite{book,arxiv,lev3} the approach to symmetry at the quantum level should be the opposite. The fact is that background is a purely classical concept. In quantum theory, each physical quantity must have a corresponding operator, but there are no operators for coordinates $x$ of the background. Therefore, the approach to symmetry at the quantum level should be as follows. Each system is described by a set of basic operators and symmetry is determined by how these operators commute with each other. For example, by definition,
dS symmetry should not
involve the fact that the dS group is the group of motions of dS space.
Instead, {\it the definition} is that the operators $M^{ab}$ ($a,b=0,1,2,3,4$, $M^{ab}=-M^{ba}$)
describing the system under consideration satisfy the
commutation relations
\begin{equation}
	[M^{ab},M^{cd}]=-i (\eta^{ac}M^{bd}+\eta^{bd}M^{ac}-
	\eta^{ad}M^{bc}-\eta^{bc}M^{ad})
	\label{CR}
\end{equation}
where $\eta^{ab}$ is the diagonal tensor such that
$\eta^{00}=-\eta^{11}=-\eta^{22}=-\eta^{33}=-\eta^{44}=1$.
The {\it definition} of AdS symmetry is given by the same equations
but $\eta^{44}=1$.

The concepts ($kg,m,s$) come from classical theory, so these concepts should not exist in quantum theory. In particular, quantum theory should not contain the parameters
($c,\hbar,R$) if $c$ in understood as the speed of light in $m/s$, $\hbar$ is understood as the Planck constant in $kg\cdot m^2/s$ and $R$ is understood as the radius of dS or AdS space in meters. With such a treatment of ($c,\hbar,R$), these parameters may be different at
different stages of the universe. However, as argued by Dyson in his famous paper \cite{Dyson}, in quantum theory,
($c,\hbar,R$) can be treated as contraction parameters from RQT to NQT, from QT to CT and from 
dSQT or AdSQT to RQT, respectively. Then the parameters ($c,\hbar,R$) can be identified with
their respected classical values in semiclassical approximation. For the first time, the concept of contraction has been
discussed by Inonu and Wigner \cite{IW}.

As argued by Dyson \cite{Dyson} (see also \cite{book,arxiv,lev3}):
\begin{itemize}
\item i): NQT is a degenerate case of RQT in the formal limit $c\to\infty$;
\item ii) CT is a degenerate case of QT in the formal limit $\hbar\to 0$;
\item iii) RQT is a degenerate case of dSQT and AdSQT in the formal limit $R\to\infty$.
\end{itemize}
In the literature, those properties are usually discussed from physical considerations.
But, as shown in Sec. 1.3 of \cite{book,arxiv}, those properties can be proved
purely mathematically taking into account {\bf Definition} and the fact that symmetry at
the quantum level is defined by the corresponding representation of the symmetry algebra.

The above facts prove that $R$ is fundamental to the same extent as $\hbar$ and $c$ (see also \cite{book,arxiv,asymm} for details). 
By analogy with the fact that $c$ must be finite, $R$ must be finite too: the formal case $R=\infty$ corresponds to the 
situation when the dS and AdS algebras do not exist because they become the Poincare algebra.
{\it At the quantum level, $R$ is only the parameter of contraction from dS or AdS algebras to the Poincare one
and has nothing to do with the radius of the dS or AdS space}. As shown in \cite{asymm}, the result for the cosmological acceleration obtained in semiclassical approximation to dSQT without any geometry is the same as in GR when the radius of the dS space equals $R$.

The properties i)-iii) have been proved in SQT based on complex numbers. The problem arises
how to generalize these results to the case of FQT. In this theory, the space of states
is a linear space over the ring $R_{p^2}$ or the field $F_{p^2}$ which contain $p^2$ elements. 
Any element of $R_{p^2}$ can be represented as $a+bi$ where $a,b\in R_p$ and $i$ is a formal element such that $i^2=-1$. Then the definition of addition, subtraction and multiplication in $R_{p^2}$ is obvious and $R_{p^2}$ is a ring regardless whether $p$ is prime or not. 

However, $F_{p^2}$ can be a field only if $p$ is prime but this condition is not sufficient.
By analogy with the field of complex
numbers, one could define division as $(a+bi)^{-1}=(a-ib)/(a^2+b^2)$.
This definition can be meaningful only if $a^2+b^2\neq 0$ in $F_p$
for any $a,b\in F_p$ i.e., $a^2+b^2$ is not divisible by $p$.
Therefore the definition is meaningful only if $p$ {\it cannot}
be represented as a sum of two squares. For example, $F_{p^2}$ can be defined
as $F_p+iF_p$ if $p=7$ but cannot be defined in this way if $p=5$.

We will not consider the case $p=2$ and therefore $p$ is necessarily odd.
Then we have two possibilities: the value of $p\,\,(mod \,4)$ 
is either 1 or 3. The known result of number theory
\cite{VdW,Ireland,Davenport} is that a prime number $p$ can be
represented as a sum of two squares only in the former case
and cannot in the latter one. Therefore $F_{p^2}=F_p+iF_p$ only if
$p\,(mod \,4)\,=\,3$. Nevertheless, as shown in standard textbooks \cite{VdW,Ireland,Davenport}, quadratic extensions of $F_p$ exist also in the case 
$p=1 (mod\,\,4)$.

Every quadratic finite ring or field has only one nontrivial automorphism $^*$. If $R_{p^2}=R_p+iR_p$ or 
$F_{p^2}=F_p+iF_p$, this automorphism is the complex conjugation: $(a+bi)^*=(a-bi)$, but, as shown in standard textbooks (e.g., in \cite{VdW,Ireland,Davenport}), the automorphism of
$F_{p^2}$ can also be defined if $p=1\,\, (mod\,\,4)$.

In spaces over $R_{p^2}$ or $F_{p^2}$ one can formally define a scalar product $(y,x)$ for
the elements $x,y$ belonging to those spaces such that $(y,\lambda x)=\lambda (y,x)$ and
$(\lambda y,x)=\lambda^* (y,x)$ where $\lambda\in R_{p^2}$ or $\lambda\in F_{p^2}$,
respectively.

In SQT, operators
$A$ of physical quantities act in Hilbert spaces ${\cal H}$ supplied by a scalar product (...,...), and these operators
are selfadjoint: $(Ax,y)=(y,Ax)\,\, \forall x,y\in {\cal H}$ belonging to the domain of $A$. In particular, the operators
in Eqs. (\ref{CR}) are selfadjoint. By analogy, in FQT linear operators
$A$ of physical quantities act in spaces over $R_{p^2}$ or $F_{p^2}$ and formally such
operators can be called selfadjoint if $(Ax,y)=(y,Ax)$ for all elements $x,y$ belonging
to such spaces.

In SM, scalar products in Hilbert spaces have a property $(x,x)>0$ if $x\neq 0$,
and in SQT this property has a known probabilistic interpretation. The physical meaning of
probability is such that it is defined by an infinite number of experiments. In nature there can be no infinite number of experiments and so the concept of probability
is based on an idealization. However, as explained in Sec. \ref{finmath}, in FM the concepts
of $>$ and $<$ have a limited meaning. For example, if $e_1,e_2,,,e_n$ are elements of the
basis in a space over $R_{p^2}$ such that $(e_j,e_k)=0$ if $j\neq k$, and $a_1,a_2,...a_n$
are elements of $R_{p^2}$ then 
\begin{equation}
(a_1e_1+...a_ne_n,a_1e_n+...a_ne_n)=a_1a_1^*(e_1,e_1)+...a_na_n^*(e_n,e_n)
\label{corr1}
\end{equation}
In SM, the analogous expression will be always positive but, since in FM the operations
are performed modulo $p$, this expression may be even ''negative'',
even if all the quantities $a_ja_j^*$ and $(e_j,e_j)$ are ''positive''. Therefore, in
FQT the probabilistic interpretation has only a limited meaning when not only
$f(a_ja_j^*)>0$ and $f((e_j,e_j))>0$ $\forall j$ but also
\begin{equation}
f(a_1a_1^*(e_1,e_1)+...a_na_n^*(e_n,e_n))>0
\label{corr2}
\end{equation} 

It is clear that only those quantum theories over SM can be generalized to theories over FM where all physical
quantities are dimensionless and discrete. As shown in \cite{book, arxiv}, among the theories considered in this section, only in dSQT and AdSQT all physical quantities are dimensionless and those theories are the most general. 

In SQT, IRs of the algebras in Eq. (\ref{CR}) when the operators in these expressions are
selfadjoint, are described in a wide literature. All such IRs are infinite-dimensional.
Representations in spaces over a ring or field of nonzero characteristic are called
modular representations. According to the Zassenhaus theorem (see e.g., \cite{Zassenhaus,Jantzen}), all modular IRs are finite-dimensional. In \cite{lev1,lev2}
we constructed modular IRs of the algebras defined by Eq. (\ref{CR}).

In SQT, all Hilbert spaces are separable, i.e., they contain a countable dense subset. As
shown in standard textbooks on Hilbert spaces (see e.g. \cite{Fomin}), 
a Hilbert space is separable if and only if it admits a countable orthonormal basis
$(e_1,e_2,...e_n,...)$. It is always possible to choose a basis such that the norm of each $e_j$ is an integer. The elements of such spaces can be denoted as 
$(c_1,c_2,...c_n,...)$ where all the coordinates $c_j$ are complex numbers. The known result 
of the theory of Hilbert spaces is that the set of all points $(c_1,c_2,...)$ with only
finitely many nonzero coordinates, each a rational number, is dense in the separable
Hilbert space (see e.g., \cite{Fomin}). This implies that {\it with any desired accuracy}
each element of the Hilbert space can be approximated by a finite linear combination
\begin{equation}
x=\sum_{j=1}^n c_je_j
\label{lin}
\end{equation}
where $c_j=a_j+ib_j$ and all the numbers $(a_j,b_j) \,\,(j=1,2,....n)$ are rational. 

The next observation is that spaces in quantum theory are projective, i.e., for any complex
number $c\neq 0$ the elements $x$ and $cx$ describe the same state. The meaning of this statement is that not the probability itself but ratios of different probabilities have
a physical meaning. As a consequence, both parts of Eq. (\ref{lin}) can be multiplied by a 
common denominator of all the nonzero numbers $a_j$ and $b_j$. As a consequence,

{\it Statement 2: Each element of a separable Hilbert space can be approximated with any desired accuracy by a finite linear combination (\ref{lin}) where all the numbers $a_j$ and $b_j$ are integers, i.e., belong to $Z$.}

{\bf The important consequence for understanding standard quantum theory is that in this theory there is a large excess of states: although formally the theory involves Hilbert spaces
of states $(c_1,c_2,...c_n,...)$ where all the $c_j$ are arbitrary complex numbers and
the only limitation is the condition $\sum_{j=1}^{\infty}|c_j|^2 <\infty$, for describing experiments with any desired accuracy it suffices to involve only states where only
a finite number of the coefficients $c_j=a_j+ib_j$ are non-zero and all the numbers
$(a_j,b_j)$ are integers.}

Now a problem arises how to use {\bf Definition} for proving that FQT is more general (fundamental) than SQT and the latter is a degenerate case of the former in the formal
limit $p\to\infty$. According to this {\bf Definition}, the proof should consist of
proving two statements:
\begin{itemize}
\item A) There exists a value of $p=p_0$ such that any result of SQT can be obtained in FQT for all $p\geq p_0$;
\item B) There exist phenomena which FQT can describe while SQT cannot. 
\end{itemize}

Let us first consider property A). 

In SQT, states of a system are described by Eq. (\ref{lin})
where the $e_j$ are elements of a basis in a Hilbert space and the $c_j=a_j+ib_j$ are complex numbers. At the same time, in FQT states of a system are also described by Eq. (\ref{lin}) but now the $e_j$ are elements of a basis in a space over $R_{p^2}$ and $c_j=a_j+ib_j$ where the elements $a_j$ and $b_j$ belong to $R_p$. As explained above, in SQT it is always possible to find the elements $e_j$ such that their norms are integers and, as noted
in {\it Statement 2}, it suffices to consider such states (\ref{lin}) that only finite numbers
of the $a_j$ and $b_j$ are non-zero integers, i.e., they are elements of $Z$. 

As noted above, it follows from i)-iii) that among quantum theories in which the symmetry algebras are ten-parameter, dSQT and AdSQT are the most general. While in RQT there are operators having dimensions expressed in terms of $(kg,m,s)$ and containing a continuous spectrum, in dSQT and AdSQT, all the operators in Eqs. (\ref{CR}) are dimensionless and, as shown in \cite{lev1,lev2}, it is possible to choose bases in which they   have only a discrete spectrum, i.e., the spectrum belonging to $Z$. 

Then, as follows from {\it Statement 1} in Sec. \ref{finmath}, if Eqs. (\ref{corr1}) and (\ref{corr2}) are satisfied at
some $p=p_0$, they will also be satisfied at all $p>p_0$. Therefore, if at some $p=p_0$, FQT gives the same results as SQT then the same will take place at all $p>p_0$, i.e., property A) is satisfied.

The property B) will be demonstrated in Sec. \ref{examples}. 

\section{Why finite mathematics is more natural than classical one}
\label{FQT}

The belief that SM is the most fundamental mathematics arose after Newton and Leibniz proposed the theory of infinitesimals more than 300 years ago. This belief was in the spirit of existing ideas that when people did not know about the existence of elementary particles, they believed that any object could be divided into arbitrarily large number of arbitrarily small parts. 
However, the very fact of the existence of elementary particles (which cannot be divided into parts) indicates that in nature there are no infinitesimals and continuity. Therefore, theories involving these concepts (including standard geometry), at best, can only be a good approximation when the discrete nature of matter is not taken into account.

It seems
unnatural that SQT involves SM with differential equations and infinitesimals. Even the name "quantum theory" reflects a belief that nature is quantized, i.e., discrete, and this name has arisen because in QT some quantities have discrete spectrum (e.g., the spectrum of the angular momentum operator, the energy spectrum of the hydrogen atom etc.). But this discrete spectrum has appeared in the framework of SM.

As a rule, physicists agree that in nature there are no infinitesimals. They say that, for example, $dx/dt$ should be understood as
$\Delta x/\Delta t$ where $\Delta x$ and $\Delta t$ are small but not infinitesimal. I ask them: but you work
with $dx/dt$, not $\Delta x/\Delta t$. They reply that since mathematics with derivatives works well then there is no need to philosophize and develop something else (and they are not familiar with FM). So, people invented
continuity and infinitesimals which do not exist in nature, created problems for themselves and now apply titanic efforts for solving those problems.

The founders of QT and scientists who essentially contributed to it were highly
educated. But they used only SM, and even now FM is not a
part of standard education for physicists. The development of QFT has shown that the theory 
contains anomalies and divergences. Most physicists considering those problems, work in the framework of SM and do not acknowledge that they arise just because this mathematics is applied.

Several famous physicists (e.g., the Nobel Prize laureates Gross, Nambu and Schwinger) discussed approaches when 
QT involves FM (see e.g., \cite{Nambu}). A detailed discussion of these approaches is
given in the book \cite{Vourdas} where they are characterized as hybrid quantum systems. The reason is that here
momenta and coordinates belong to a finite ring or field but wave functions are elements of standard Hilbert spaces.
Then the problem of foundation of QT is related to the problem of foundation of SM.
On the other hand, in \cite{book,arxiv,lev3,lev1,lev2}, we have proposed an approach called finite quantum theory (FQT) where not only physical quantities but also wave functions involve finite rings or fields.

In view of this discussion, a problem arises whether it is justified to use mathematics with infinitesimals for describing nature in which infinitesimals do not exist. Although 
SM describes many physical phenomena with a very high accuracy, a problem arises whether there are phenomena which cannot be correctly described by mathematics involving infinitesimals.

Some facts of SM seem to be unnatural. For example,
$tg(x)$ is one-to-one reflection of $(-\pi/2,\pi/2)$ onto $(-\infty,\infty)$, i.e., the impression might arise that
the both intervals have the same numbers of elements although the first interval is a nontrivial part of the second one. However, Hilbert said: "No one shall expel us from the paradise that Cantor has created for us”. 

From the point of view of Hilbert's approach (see Sec. \ref{goal}) it is not important whether
some statements of SM are natural or not, since the goal of the approach is to find a complete and consistent set of axioms.
In the framework of this approach, the problem of foundation of SM has been investigated by many great mathematicians
(e.g., Cantor, Fraenkel, G\"{o}del, Hilbert, Kronecker, Russell, Zermelo and others). Their philosophy was based on
macroscopic experience in which the concepts of infinitesimals, continuity and standard division are natural. However, as noted above, those concepts contradict the existence of elementary particles and are not natural 
in QT. The illusion of continuity arises when one neglects the discrete structure of matter.

The existence of foundational problems in Hilbert's approach follows, in particular, from
Gödel’s incompleteness theorems which state that no system of axioms can ensure that all facts about natural numbers can be proved, and the system of axioms in SM cannot demonstrate
its own consistency. The theorems are written in highly technical terms of mathematical logics. 
As already noted, in this paper we do not consider Hilbert's approach to mathematics.
However, simple arguments in \cite{book,arxiv} show that, if mathematics is treated as a tool for describing nature,
then foundational problems of SM
follow from simple arguments described below.  

In the 20s of the 20th century, the Viennese circle of philosophers
developed an approach called logical positivism which contains
verification principle:  {\it A proposition is only cognitively meaningful if it can be definitively and conclusively 
determined to be either true or false} \cite{verif1,verif2}. However, this principle does not work if SM is treated as a tool for describing nature. 
For example, in Hilbert's approach one of axioms is that $a+b=b+a$ for all natural numbers $a$ and $b$,
and a question whether this is true or false does not arise. However, if mathematics is treated as a tool
for describing nature, it cannot be determined whether this statement is true or false. 

As noted by Grayling \cite{Grayling}, {\it "The general laws of science are not, even in principle, verifiable, if verifying means 
furnishing conclusive proof of their truth. They can be strongly supported by repeated experiments and 
accumulated evidence but they cannot be verified completely"}. So, from the point of view of
SM and physics, verification principle is  
too strong. 

Popper proposed the 
concept of falsificationism \cite{Popper}: {\it If no cases where a claim is false can be found, then 
the hypothesis 
is accepted as provisionally true}. In particular, the statement that 
$a+b=b+a$ for all natural numbers $a$ and $b$ can be treated as provisionally true until one has found
some numbers $a$ and $b$ for which $a+b\neq b+a$.

According to the philosophy of quantum theory, there should be 
no statements
accepted without proof and based on belief in their correctness (i.e., axioms).
The theory should contain only those statements that can be verified, where by "verified" physicists mean 
an experiment involving only a finite number of steps. This philosophy is the result of the fact that quantum theory describes phenomena which, from the point of view of “common sense”, seem meaningless but they have been experimentally verified.
So, the philosophy of QT is similar to 
verificationism, not falsificationism. Note that Popper was a strong opponent of 
QT and supported Einstein in his dispute with Bohr.

From the point of view of verificationism and the philosophy of QT, SM is not well 
defined not only because it contains an infinite number of numbers. Consider, for example, whether the
rules of standard arithmetic can be justified.

We can verify that 100+100=200 and 1000+1000=2000, but can we verify that, say $10^{100000}+10^{100000}=2\cdot 10^{100000}$? One might think that this is obvious, and in Hilbert's approach this follows from main axioms. 
But, if mathematics is treated as a tool for describing nature then this is only a belief based on extrapolating our everyday experience to numbers where it is not clear whether the experience still works. 

In Sec. \ref{SR} we discussed that our life experience works well at speeds that are much less than $c$, but this experience cannot be extrapolated to situations where speeds are comparable to $c$. Likewise, our experience with the numbers we deal with in everyday life cannot be extrapolated to situations where the numbers are much greater.

According to verificationism and principles of quantum theory, the statement that $10^{100000}+10^{100000}=2\cdot 10^{100000}$ is true or false depends on whether this statement can be verified. Is there a computer which can verify this statement? Any computing device can operate only with a finite number of resources and can perform calculations only modulo some number $p$. If our universe contains only a finite number of elementary particles, then in principle it is not possible to verify that standard rules of arithmetic are valid for any numbers.

That is why the statements in Eq. (\ref{standard}) are ambiguous because they do not contain information on 
the computing device which verifies those statements. For example, let us pose a problem whether 100+200 equals 300. If our computing devise is such that $p=400$ then the experiment will confirm this while if $p=250$ then we will get that 100+200=50. 

{\it So, the statements that 100+200=300 and even that $2\cdot 2=4$
are ambiguous because they do not contain information on how they should be verified.} 
On the other hand, the statements
$$100+200=300\,(mod\, 400),\,\, 100+200=50\,(mod\, 250),$$
$$2\cdot 2=4\,(mod\, 5),\,\, 2\cdot 2=2\,(mod\, 2)$$
are well defined because they do contain such an information.
So only operations modulo a number are well defined.

We believe the following 
observation is very important. Although SM is a part of our everyday life, people typically do not realize that {\it standard
operations with natural numbers are implicitly treated as limits of operations modulo $p$ 
when $p\to\infty$}. For example, if $(a,b,c,p)$ are natural numbers then  Eqs. (1) 
are implicitly treated as
$$\lim_{p\to\infty} [(a+b)\,\,(mod\,\,p)]=c,\,\, \lim_{p\to\infty} [(a\cdot b)\,\,(mod\,\,p)]=c,\,\,etc.$$
 
As a rule, every limit in mathematics is thoroughly investigated but, in the
case of standard operations with natural numbers, it is not even mentioned that those
operations are limits of operations modulo $p$. In real life such limits even 
might not exist if, for example, the universe contains a finite number of elementary particles.
 
{\it So, we see that the question of what 100+200 is equal to is not a question of what some theory says, but a question of how an experiment will be set up to test what this value is equal to. In one experiment the result may be 300, in another 50 and there is no theory that says that one experiment is more preferable than another.}

Now let's discuss the question of what $p$ can be equal to in the theory describing modern physics. Recently, an increasing number of works have appeared that say that the universe works like a computer (see, for example, \cite{Wolfram}). From this point of view, the value of $p$ is determined by the state of the universe at a given stage. And, since the state of the universe is changing, it is natural to expect that the number $p$ describing physics at different stages of the evolution of the universe will be different at different stages. Therefore, by analogy with the discussion of what 100+200 is equal to, we can say that $p$ is not a number that is determined by some fundamental theory, but a number that depends on the state of the universe at a given stage.

The problem of time is one of the most fundamental problems of quantum
theory.  Every physical quantity should be described by a selfadjoined operator but, as noted by Pauli, the existence of the time
operator is a problem (see e.g., the discussion in \cite{book,arxiv}). One of the principles of physics is that the
definition of a physical quantity is a description how this quantity should be measured, and {\it it is not correct to say that
some quantity exists but cannot be measured}. The present definition of a second is the time during which 
9,192,631,770 transitions in a cesium-133 atom occur. The time cannot be measured with  absolute accuracy
because the number of transitions is finite. Then one second is defined with the 
accuracy $10^{-15}s$, and \cite{time} describes efforts to measure time with the accuracy $10^{-19}s$.
However, a problem arises how to define time in early stages of the universe when atoms did not exist. So, treating time $t$ as a continuous quantity can be only an approximation which works at some conditions.
In \cite{book,arxiv} we discussed a conjecture that standard classical time $t$ manifests itself
because the value of $p$ changes, i.e., $t$ is a function of $p$. We do not say that $p$ changes
over time because classical time $t$ cannot be present in quantum theory; we say that
we feel changing time because $p$ changes. As shown in \cite{asymm} (see also the subsequent section), with such an approach, the known problem of baryon asymmetry of the universe 
does not arise.

\begin{sloppypar}
\section{Examples when finite mathematics can solve problems which standard mathematics cannot}
\label{examples}
\end{sloppypar}

As noted in Sec. \ref{FQTvsSQT}, for proving that FQT is more general (fundamental) than
SQT it is necessary to prove the properties A) and B) described at the end of this section.
Property A) has already been demonstrated at the end of Sec. \ref{FQTvsSQT}. Property B) means that there are phenomena that FQT can explain, but SQT cannot. 
In \cite{book,arxiv} we discussed phenomena
where it is important that $p$ is finite. They cannot be described in SQT, by analogy with the fact that NM cannot describe cases where
it is important that $c$ is finite. Below we describe several such phenomena.

{\bf Example 1: gravity}. The Newton gravitational law cannot be derived in QFT because the theory is not renormalizable. But the law can be derived from FQT in semiclassical approximation \cite{book,arxiv}. Then the gravitational constant $G$ is not taken from the outside but depends on
$p$ as $1/ln(p)$.
By comparing this result with the experimental value, one gets that $ln(p)$ is of the order of $10^{80}$ or more, and therefore $p$ is a huge number of the order of $exp(10^{80})$ or more. 
One might think that since $p$ is so huge then in practice $p$ can be treated as an infinite number. However, since $ln(p)$ is "only" of the order of $10^{80}$, gravity is observable.
In the formal limit $p\to\infty$, $G$ becomes zero and gravity disappears. Therefore, in our approach, gravity is
a consequence of finiteness of nature.

{\bf Example 2: Dirac vacuum energy problem}. In quantum electrodynamics, the vacuum energy should be zero, but in QFT the sum for this energy diverges, and this problem was posed by Dirac. To get the zero value, the artificial requirement that the operators should be
written in the normal order is imposed, but this requirement does not follow from the construction of the theory.
In Sec. 8.8 of \cite{book,arxiv}, I take the standard expression for this sum and explicitly calculate it in FM without any assumptions. Then, since the calculations are modulo $p$, I get zero as it should be.

{\bf Example 3: equality of masses of particles and their antiparticles}. This is an example
demonstrating the power of finite mathematics. 
A discussion in \cite{book,arxiv,lev3} and Sec. \ref{FQTvsSQT} shows that in QT, an elementary particle and its antiparticle should be considered only from the point of view of IRs of the symmetry algebra. In SQT, the algebras are such that their IRs contain either only positive or only negative energies. In the first case the objects are called particles and in the second one – antiparticles. Then the energies of antiparticles become positive after second quantization. 

In QFT, the spectrum of positive energies contains the values $(m_1, m_1+1, m_1+2,\cdots\infty)$, and for negative energies --- the values $(-m_2, -m_2-1, -m_2-2,\cdots -\infty)$, where $m_1>0,\, m_2>0$, $m_1$ is called the mass of a particle and $m_2$ is called the mass of the corresponding antiparticle. Experimentally $m_1=m_2$ but in QFT, IRs with positive and negative energies are fully independent of each other. 
It is claimed that $m_1=m_2$ because local covariant equations 
are CPT invariant. However, as explained in
\cite{book,arxiv,lev3}, the argument $x$ in local quantized fields does not have a physical meaning because it is not associated with any operator. 
So, in fact, SQT cannot explain why $m_1=m_2$.

Consider now what happens in FQT. For definiteness, we consider the case when
$p$ is odd, and the case when $p$ is even can be considered analogously. One starts constructing the IR with the value $m_1$, and, by acting on the states by raising operators, one gets the values $m_1+1, m_1+2,\cdots$. However, now we are moving not along the $x$ axis but along the circle in Figure 1. When we reach the value $(p-1)/2$, the next value is $–(p-1)/2$, i.e., one can say that by adding 1 to a large positive number $(p-1)/2$ one gets a large negative number $–(p-1)/2$. By continuing this process, one gets the numbers $-(p-1)/2+1=-(p-3)/2$, $-(p-3)/2+1=-(p-5)/2$ etc. The explicit calculation \cite{book,arxiv} shows that the procedure ends when the value $-m_1$ is reached.

Therefore, FM gives a clear proof that $m_1=m_2$ and shows that, instead of two independent IRs in SM, one gets only one IR describing both, a particle, and its antiparticle. The case described by SM is degenerate because, in the formal limit $p\to\infty$, one IR in FM splits into two IRs in SM. So, when $p\to\infty$ we get symmetry breaking. This example
shows that standard concept of particle-antiparticle is only approximate and is approximately valid only when $p$ is very large. Therefore, constructing complete QT based on FM should be based on new principles. 

{\bf Example 4: the problem of baryon asymmetry of the universe}.  
Modern cosmological theories state that the numbers of baryons and antibaryons in the early stages of the universe were the same. Then, since the baryon number is the conserved quantum number, those numbers should be the same at the present stage. However, at this stage the number of baryons is much greater than the number of antibaryons.

For understanding this problem, one should understand the concept of particle-antiparticle.
In SQT this concept takes place because IRs describing particles and antiparticles are
such that energies in them can be either only positive or only negative but cannot have both signs. However, as explained
in {\bf Example 3}, IRs in FQT necessarily contain both, positive and negative energies, and in the formal limit
$p\to\infty$, one IR in FQT splits into two IRs in SQT with positive and negative energies.

As noted above, the number $p$ is different at different stages of the universe. As noted in {\bf Example 1}, at the present stage of the universe this number is huge, and therefore
the concepts of particles and antiparticles have a physical meaning. However, arguments given in \cite{book,arxiv} indicate that in early stages of the universe the value of $p$ was much less than now.
Then each object described by IR, is a superposition of a particle and antiparticle (in SQT such a
situation is prohibited), and the electric
charge and baryon quantum number are not conserved. Therefore, in early stages of the universe, SQT
does not work, and the statement that at such stages the numbers of baryons and antibaryons were the same,
does not have a physical meaning. Therefore, the problem of baryon asymmetry of the universe does not arise.

{\bf Example 5:} As argued in Sec. 6.8 of \cite{book,arxiv}, the ultimate QT will
be based on a ring, not a field, i.e., only addition, subtraction and multiplication are
fundamental mathematical operations, while division is not.

The above examples demonstrate that there are phenomena which can be explained only in FQT because for them it is important that $p$ is finite and not infinitely large. So, FQT is
more general (fundamental) than SQT. Here we have an analogy with the case that SR can explain phenomena where $c$ is finite while NM cannot explain such phenomena. 

\section{Answers to arguments (a-c) in Sec. \ref{goal}}
\label{answers}

To get rid of divergences, physicists usually do the following. In integrals over the absolute values of momenta, the upper limit of integration is taken not $\infty$ as it should be, but a value $L$ called the Pauli-Villars cutoff. Then all integrals formally become finite, but they depend on the nonphysical very large quantity $L$. In renormalizable theories, various contributions to the S-matrix can be arranged in such a way that the contributions with $L$ cancel, but in non-renormalizable theories it is not possible to get rid of L.

The idea of argument a) is such that, by analogy with SQT, where there are divergent integrals that are cut off by the value of $L$, in FQT there are formally no divergences but there are quantities depending on the enormous value $p$.
However, this analogy doesn't work for several reasons. 

In Sec. \ref{SR} we noted that, from our experience in NM, we think that some of the arguments are based on common sense. But these arguments only work at speeds which are much less than $c$ and often fail at speeds comparable to $c$. Likewise, some arguments which, from our experience in SM, seem to come from common sense, usually work in FM only for numbers much less than $p$, and often fail for numbers comparable to $p$.

As noted in Sec. \ref{finmath}, in FM there are no strict concepts of positive and negative and the concepts of $>$ and $<$. These concepts approximately work for numbers that are much less than $p$ and are in some neighborhood of zero on Figure 1.

In SM, when we add two positive numbers, we always get a positive number that is greater than the original numbers. However, since in FM calculations are carried out modulo $p$, situations are possible when we add two ''positive'' numbers and get a ''negative'' number. For example, in finite mathematics, $(p-1)/2 +1=-(p-1)/2$, i.e.,
adding two numbers which in Figure 1 are in the right half-plane, we get a number that in this figure is in the left half-plane.

In {\bf Example 2} in Sec. \ref{examples}, we describe an example when in SQT, as a result of adding many positive values, a divergent expression is obtained, while in FQT the result is 0 because the calculations are carried out modulo $p$. Thus, argument a) does not always work in FQT.

The argument b) is unacceptable even because the theory with adeles is not finite
and therefore automatically has foundational problems. The arguments b) and c) (that
it is not clear from what considerations $p$ is chosen) is not a refutation of FQT for the following reason. As explained in Sec. \ref{FQT}, the value of $p$ is not a fundamental parameter that follows from some theory: this value is determined by the state of the universe at the given stage of its development, and at different stages the values of $p$ are different.

To conclude this section, we note the following. One of the objections to FQT is that the authors of these objections interpret $p$ as the greatest possible number in nature and invoke the argument attributed to Euclid that there can be no greatest number in nature because if $p$ is such a number then $(p+1)>p$. 
Similarly, one can say that $c$ cannot be the greatest possible speed because $1.01c > c$. As explained above, these arguments arise because our experience at speeds which are much less than $c$ and numbers which are much less than $p$ 
is extrapolated to situations where speeds are comparable to $c$ or numbers are comparable to $p$. 

\section{Conclusion}

The goal of this paper is to explain at the simplest possible level
why FM is more general (fundamental) than SM. As noted in Sec. \ref{FQT}, the belief of most mathematicians and physicists that SM is the most fundamental arose for historical reasons. However, as explained in Sec. \ref{finmath}, simple mathematical arguments show that SM (involving the concept of infinities) is a degenerate case of FM: SM arises from FM in the degenerate case when operations modulo a number are discarded.

We call FQT a quantum theory based on FM. It is determined by a parameter $p$ which is the characteristic of the ring in finite mathematics describing physics. We note that 
in FQT there are no infinities and that is why divergences are absent in principle.
Probabilistic interpretation of FQT is only approximate: it applies only to states described by numbers which are much less than $p$.

In Sec. \ref{FQT} we give arguments that $p$ is not a fundamental quantity that is determined by some theory, but depends on the state of the universe at a given stage. Therefore, $p$ is different at different stages of the universe. 

The question of why $p$ is this and not another is similar to the question of why
the values of ($c,\hbar$,R) are such and not others. As explained in \cite{book,arxiv,asymm}, currently they are such simply because people want to measure
$c$ in $m/s$, $\hbar$ in $kg\cdot m^2/s$ and $R$ in meters, and it is natural to expect that these values at different stages of the universe are different.

As noted in
Sec. \ref{examples}, at the present stage of the universe, 
$p$ is an enormous quantity of the order of $exp(10^{80})$. Therefore, at present, SM almost always works with very high accuracy.
At the same time, in \cite{book,arxiv,asymm} and Sec. \ref{examples} we argue that
in early stages of the universe, $p$ was much less than now. Therefore, at these stages, the finitude of mathematics played a much greater role than it does now. As a result, the problem of baryon asymmetry of the universe does not arise.  

The famous Kronecker's expression is: "God made the natural numbers, 
all else is the work of man”. However, in view of the above discussion, I propose to reformulate this expression as:
"God made only finite sets of natural numbers, all else is the work of man”. For illustration, consider a case when some experiment is conducted $N$ times, the first event happens $n_1$ times, the second one --- $n_2$ times etc.
such that $n_1+n_2+...=N$. Then the experiment is fully described by a finite set of natural numbers. But
people introduce rational numbers $w_i=w_i(N)=n_i/N$, introduce the concept of limit and define probabilities as
limits of the quantities $w_i(N)$ when $N\to\infty$.

The above discussion shows
that FM is not only more general (fundamental) than SM
but, in addition, in FM there are no foundational problems because every statement can be explicitly verified
by a finite number of steps. The conclusion from the above consideration can be formulated as:

{\bf Mathematics describing nature at the most fundamental level involves only a finite number of numbers, while the concepts of limit, 
infinitesimals and continuity are needed only in calculations describing nature approximately.}

{\bf Acknowledgements:} I am grateful to Justin Clarke-Doane who in numerous letters explained to me various philosophical issues of mathematics.
I am also grateful to Vladimir Karmanov, Teodor Shtilkind and the reviewers of this paper, whose comments were important in preparing the revised version of the paper. 

{\bf Funding}: This research received no external funding.

{\bf Conflicts of Interest}: The author declares no conflicts of interest.

\end{document}